\documentclass[conference]{IEEEtran}
\IEEEoverridecommandlockouts
\usepackage{cite}
\usepackage{amsmath,amssymb,amsfonts}
\usepackage{algorithmic}
\usepackage{graphicx}
\usepackage{textcomp}
\usepackage{xcolor}
\usepackage{bm} 
\def\BibTeX{{\rm B\kern-.05em{\sc i\kern-.025em b}\kern-.08em
    T\kern-.1667em\lower.7ex\hbox{E}\kern-.125emX}}
\begin{document}

\title{Unfolded Deep Neural Network (UDNN) for High Mobility Channel Estimation\\
}

\author{\IEEEauthorblockN{Yinchuan Li} 
\IEEEauthorblockA{\textit{Electrical Engineering Department} \\
\textit{Columbia University}\\
New York, USA \\
yinchuan.li.cn@gmail.com}
\and
\IEEEauthorblockN{Xiaodong Wang} 
\IEEEauthorblockA{\textit{Electrical Engineering Department} \\
\textit{Columbia University}\\
New York, USA \\
wangx@ee.columbia.edu}
\and
\IEEEauthorblockN{Robert L. Olesen}
\IEEEauthorblockA{\textit{Interdigital, Inc.} \\
New York, USA \\
robert.olesen@interdigital.com}
}

\maketitle

\begin{abstract}
High mobility channel estimation is crucial for beyond 5G (B5G) or 6G wireless communication networks. This paper is concerned with channel estimation of high mobility OFDM communication systems. First, a two-dimensional compressed sensing problem is formulated by approximately linearizing the channel as a product of an overcomplete dictionary with a sparse vector, in which the Doppler effect caused by the high mobility channel is considered. 
To solve the problem that the traditional compressed sensing algorithms have too many iterations and are time consuming, we propose an unfolded deep neural network (UDNN) as the fast solver, which is inspired by the structure of iterative shrinkage-thresholding algorithm (ISTA).
All the parameters in UDNN (e.g. nonlinear transforms, shrinkage thresholds, measurement matrices, etc.) are learned end-to-end, rather than being hand-crafted.
Experiments demonstrate that the proposed UDNN performs better than ISTA for OFDM high mobility channel estimation, while maintaining  extremely  fast computational speed.
\end{abstract}


\section{Introduction}
High mobility communication is one of the main challenges in wireless communication networks, especially for the fifth generation (5G) and beyond 5G (B5G) systems~\cite{albreem20155g,fan20165g}. Since the Doppler shift caused by the high mobility is non-negligible, for single-antenna systems, the channel estimation is a two-dimensional (2D) spectrum estimation problem, where the channel model can be represented by a matrix with the two dimensions denoting time delays and Doppler shifts, respectively. To estimate these parameters, with the {\em compressed sensing} (CS) technique~\cite{donoho2006compressed,candes2011compressed} can explore the sparsity feature of channels to improve the estimation performance~\cite{ma2017design, han2017compressed, li2019interference}.

In the past decade, extensive CS methods for spectral estimation have been developed, based on some well-studied estimation models, these methods enjoy the advantages of strong convergence and theoretical analysis in most cases. Especially the methods based on the atomic norm (AN)~\cite{tang2013compressed,yang2014exact,bhaskar2013atomic} minimization techniques have attracted considerable interest in spectral super-resolution. Compared with these continuous sparse recovery techniques, the {\em on grid} approximation based methods can have lower computational complexity, since the linearized sparsity-regularized optimization problem can be solved in an iterative fashion, for example, via the iterative shrinkage-thresholding algorithm (ISTA). However, the computational complexity of these methods is still high, and they also face the challenge of choosing and tuning parameters.

Fueled by the powerful learning ability of deep networks, several deep network-based CS estimation algorithms have been recently proposed to directly learn the inverse mapping from the CS measurement domain to the original signal domain. Compared to optimization-based algorithms, these non-iterative algorithms dramatically reduce computational complexity, while achieving impressive estimation performance. However, many network-based CS algorithms are trained as a black box, with limited insights from the CS domain. Recently, the idea of mapping the ISTA into a recurrent neural network (RNN) using its iterative properties has attracted widespread attention in the image processing field~\cite{zhang2018ista,yang2018admm,ito2019trainable,gregor2010learning}.

In this paper, to solve the high mobility channel estimation problem, we model the channel estimation as a two-dimensional compressed sensing problem with one dimension corresponding to the delay and the other corresponding to the Doppler caused by the high mobility channel. Then, to meet the real-time processing requirements, we propose an unfolded deep neural network (UDNN), which can quickly solve the compressed sensing problem by mapping the ISTA solver into a deep network. In particular, we design two parallel propagation lines in the network that are respectively for the real and imaginary parts calculation, and information in these lines is interleaved with each other to ensure that the network operations are equivalent to complex operations. In addition, UDNN includes two repetitive phases respectively corresponding to the two iterative steps in ISTA. Instead of the traditional linear transformations, nonlinear learnable and sparse transformations are adopted in UDNN, all parameters involved in UDNN (e.g. nonlinear transforms, shrinkage thresholds, measurement matrix, etc.) are learned end-to-end by using back-propagation, instead of being fixed or hand-crafted. UDNN has the advantages of fast and accurate reconstruction with well-defined interpretability. Channel estimation experiments clearly show that UDNN performs better than ISTA, while maintaining attractive computational complexity.

The remainder of this paper is organized as follows. In Section II, we present the problem model for OFDM channel estimation and its ISTA solver.
In Section III and IV, we respectively propose our UDNN and the training details.
Simulation results for channel estimation via UDNN and ISTA given in Sections V. Section VI concludes the paper.

\section{Signal Model \& Problem Formulation}

\subsection{High Mobility Signal Model}

We assume that the signal $s(t)$ is the OFDM signal that is widely adopted in contemporary wireless communication systems. The OFDM system consists of $N_d$ data sub-carriers and $N_uT = (N_d + N_p)T$ basic time units, with $N_p$ being the number of cyclic prefix (CP) carriers and $T$ being the sampling period (``sub-pulse duration''). Then, the transmitted baseband OFDM signal over $N_b$ blocks is given by
\begin{equation}
\label{eq:st}
s(t)= \sum_{n=0}^{N_b-1} \sum_{k=0}^{N_d-1} b_n(k)e^{i2\pi k \frac{t}{N_dT} } u(t - n N_uT),
\end{equation}
where $b_n(k), ~k = 0,...,N_d-1$ is the $n$-th normalized data symbol block, such that $\mathbb{E}[b_n(k)b_n(k)^*]=1$ with $(\cdot)^*$ denoting the complex conjugate operator; and 
\begin{equation}
u(t) = \left\{
   \begin{aligned}
   1,&~~t\in[-N_pT,N_dT],  \\
   0,&~~\text{otherwise}. \\
   \end{aligned}
   \right.
\end{equation}

We consider the Doppler effect caused by high mobility in the channel model. Since in 5G mm-wave systems, the considered speed for high mobility channel is in the range of 0-500 km/h, and the carrier frequency is in the range of 28-300 GHz with signal bandwidth on the order of 1 GHz~\cite{li2020multi}, such that for $\ell$-th target with complex gain $c_{\ell}$, delay $\bar\tau_{\ell}$ and Doppler shift ${\bar f}_{\ell}$ we have ${\bar f}_{\ell}N_uT \ll 1$. Hence, the phase rotation due to the Doppler shift can be approximated as constant over an OFDM symbol duration $N_uT$, i.e.,~\cite{berger2010signal}
\begin{align}
e^{i2\pi {\bar f}_{\ell} t} \approx e^{i2\pi {\bar f}_{\ell} n {N_uT}}, ~ t \in [nN_uT,(n+1)N_uT].
\end{align}

 At the receiver side, we only refer to the baseband signals by assuming that down-conversion has been performed. The CP is removed assuming that its length is no less than the maximum path delay. Furthermore, in the following, we assume that an estimate of the data symbols, $\hat b_n(k)$, is available.
Then, in the $n$-th OFDM symbol, matched filtering is performed to obtain, for $k = 0,...,N_d-1$~\cite{li2019multi},
\begin{align}
\bar y_{n}(k) &= { \sum_{\ell=1}^{J} c_{\ell} }  \sum_{q=0}^{N_d-1}\hat  b_n(q)  \frac{1}{N_dT}  \int_{nN_uT}^{nN_uT+N_d T} \underbrace{e^{i2\pi {\bar f}_{\ell} t}}_{\approx e^{i2\pi {\bar f}_{\ell} n {N_uT}}} \nonumber \\
&~~~~~~~~~~~~~~~~~~~~~~e^{i2\pi q \frac{t - {\bar\tau}_{\ell} }{N_dT} } {e^{\frac{{ - i2\pi k t}}{N_dT}}} dt + \bar w_{n}(k) \nonumber \\
 \label{eq:ynm2}
&\approx \hat b_n(k)     \sum\limits_{\ell = 1}^{J} c_{\ell} e^{i2\pi n f_{\ell}} {e^{ - i2\pi k \tau_{\ell}}} + \bar w_{n}(k),
\end{align}
where $\bar w_{n}(k)$ is a white, complex circularly symmetric Gaussian process and
\begin{align}
\label{eq:tau-DeltaM}
\tau_{\ell} = \frac{{\bar\tau}_{\ell} }{N_dT} \in [0,1),~f_{\ell} =&~ {\bar f}_{\ell}{N_uT} \in [0,1).
\end{align}

\subsection{Problem Formulation}

Let us now define ${\bm{c}} = [c_{1},c_{2},...,c_{J}]^T \in \mathbb{C}^{J \times 1}$, $\bm f = [f_{1},f_{2},...,f_{J}]^T \in \mathbb{C}^{J \times 1}$ and $\bm \tau = [\tau_{1},\tau_{2},...,\tau_{J}]^T \in \mathbb{C}^{J \times 1}$, and the steering vectors $\bm s(f) = [1,e^{i2\pi f},...,e^{i2\pi(N_b-1)f}]^T \in \mathbb{C}^{N_b \times 1}$ and $\bm d(\tau) = [1,e^{i2\pi\tau},...,e^{i2\pi(N_d-1)\tau}]^T \in \mathbb{C}^{N_d \times 1}$.
Correspondingly, the response matrices are defined as $\bm S(\bm f) = [\bm s(f_{1}), \bm s(f_{2}), ..., \bm s(f_{J})] \in \mathbb{C}^{N_b \times J}$ and $\bm D(\bm\tau) = [\bm d(\tau_{1}), \bm d(\tau_{2}), ..., \bm d(\tau_{J})] \in \mathbb{C}^{N_d\times J}$.
Then \eqref{eq:ynm2} can be written as the following matrix form
\begin{eqnarray}
\label{eq:Y2}
\bm{\bar Y} = \bm{\hat{B}}  \odot (\bm S(\bm f) {\rm{diag}}(\bm c) \bm D(\bm\tau)^{\rm{H}}) + \bm{\bar W},
\end{eqnarray}
where $\odot$ denotes the Hadamard product; ${\rm{diag}}(\bm c)$ denotes the diagonal matrix whose diagonal entries are $\bm c$; $\bm{\bar Y} \in \mathbb{C}^{N_b\times N_d}$, $\bm{\hat{B}} \in \mathbb{C}^{N_b\times N_d}$ and $\bm{\bar W} \in \mathbb{C}^{N_b\times N_d}$ are matrices whose $(n,k)$-th element are $\bar y_{n}(k)$, $\hat b_n(k)$ and $\bar w_{n}(k)$, respectively; and $(\cdot)^{\rm{H}}$ denotes the conjugate transpose.

{Further denote  $\bm{\hat b} = {\rm{vec}}(\bm{\hat{B}}) \in \mathbb{C}^{N_bN_d\times 1}$,  $\bm{ w} = {\rm{vec}}(\bm{\bar W}) \in \mathbb{C}^{N_bN_d\times 1}$ with ${\rm{vec}}(\cdot)$ being the vectorization operator and}
\begin{align}
\label{eq:nu}
\bm\phi =&~ \sum_{\ell=1}^{J}c_{\ell} \bm a(\tau_{\ell},f_{\ell}) \in \mathbb{C}^{N_bN_d\times 1}, \\
\label{eq:atau-f}
\text{with}~\bm a(\tau,f) =&~ \bm d(\tau)^*\otimes \bm s(f) \in \mathbb{C}^{N_bN_d\times 1},
\end{align}
and $\otimes$ being the Kronecker product. Then, we vectorize $\bm{\bar Y}$ in \eqref{eq:Y2} to obtain
\begin{align}
\label{eq:y1}
\bm{ y} =&~ {\rm{vec}}(\bm{\bar Y}) ={\rm{diag}}(\bm{\hat b} ) \left( \bm D(\bm\tau)^* \circ \bm S(\bm f) \right) \bm c  + \bm{ w} 
\end{align}
where $\circ$ is the Khatri-Rao product;  $\left( \bm D(\bm\tau)^* \circ \bm S(\bm f)\right) \in \mathbb{C}^{N_bN_d\times J}$ is a matrix whose $\ell$-th column has the form of $\bm d^*(\tau_{\ell})  \otimes \bm s(f_{\ell})$. 
Next, we linearize  $\left( \bm D(\bm\tau)^* \circ \bm S(\bm f) \right) \bm c $ as a product of an overcomplete dictionary $\bm \Phi$ of complex sinusoids with a sparse vector
\begin{align}
\label{eq:y2}
\bm{ y}  \approx&~ {\rm{diag}}(\bm{\hat b}) \bm \Phi \bm x  + \bm{ w},
\end{align}
where  $\bm x \in \mathbb{C}^{M \times 1}$ with $M\geq N_bN_d$  and the columns $\bm \Phi \in \mathbb{C}^{N_bN_d \times M} $ have the same 2D complex sinusoids form as $\bm d^*(\tilde\tau)  \otimes \bm s(\tilde f)$ with $\tilde\tau$ and $\tilde f$ respectively denoting the discrelized frequency grid point in delay and Doppler domain.

Denote $\bm A  = {\rm{diag}}(\bm{\hat b}) \bm \Phi \in \mathbb{C}^{N \times M}$ with $N = N_bN_d$, the LASSO formulation can be used to estimate the vectorized high mobility channel $\bm x$ from \eqref{eq:y2}
\begin{align}
\min_{\bm x}\frac{1}{2}\|\bm y-\bm A \bm x\|_2^2 + \lambda\|\bm x\|_1,
\label{eq: lasso}
\end{align}
where $\lambda$ is a weight factor. In the next section, we introduce the proposed UDNN for solving \eqref{eq: lasso} with low computational complexity.

\begin{figure*}[htbp]
\centerline{\includegraphics[width=6.0in]{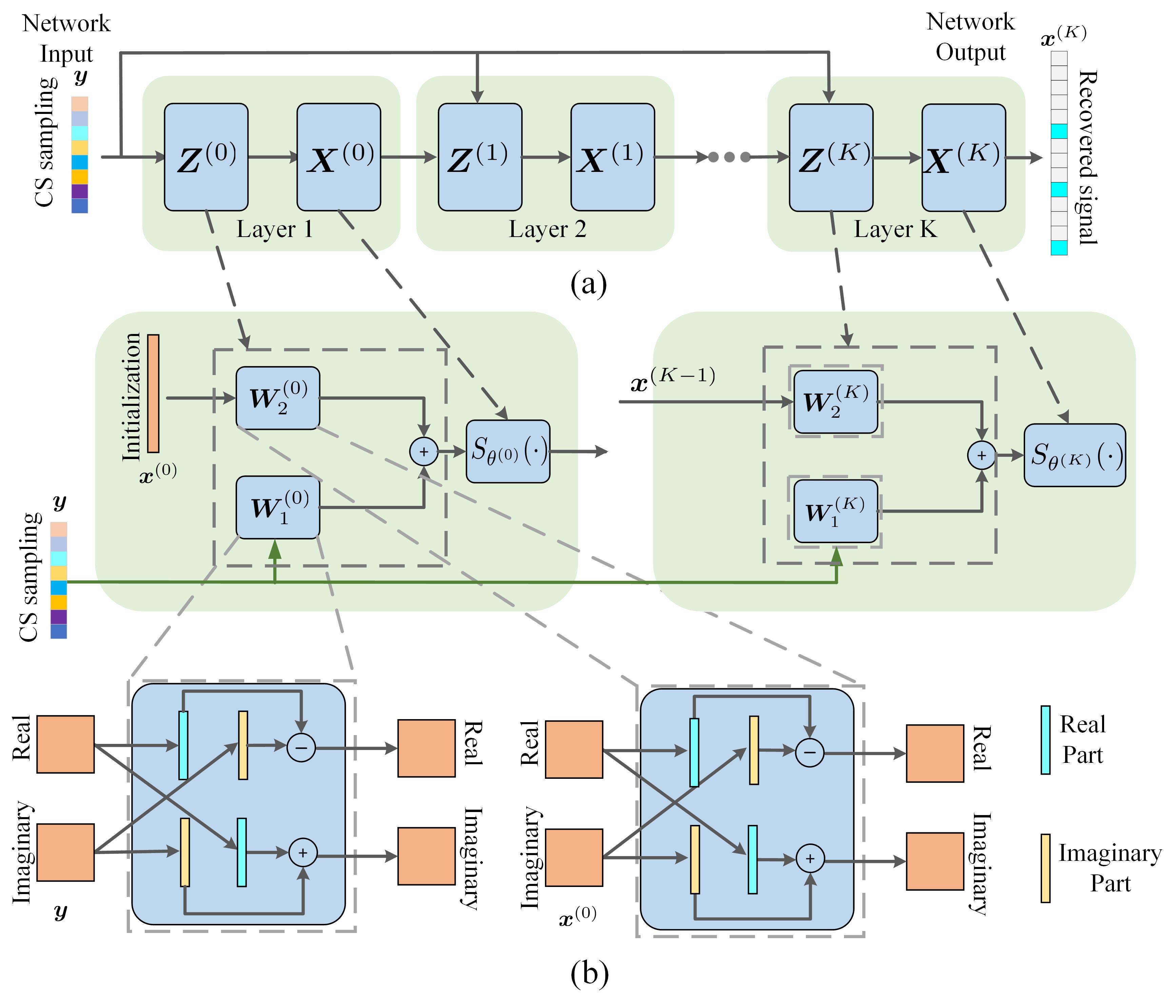}}
\caption{Diagrams of UDNN. (a) The data flow graph of UDNN. (b) Unfolded UDNN. }
\label{figure:UDNN}
\end{figure*}

%
%

\section{UDNN Framework}

\subsection{Towards UDNN}

The idea behind of the UDNN is to map an iterative algorithm for solving \eqref{eq: lasso} to a deep neural network. To start with, we present the ISTA solver here, which is widely used for solving compressed sensing problems. Specifically, ISTA solves \eqref{eq: lasso} by iterating the following two steps:
\begin{equation}
        \left\{ \begin{array}{l}
                {{\cal{Z}}^{(k)}}:\bm{z}^{(k)}=\bm{x}^{(k-1)}-\rho \bm{A}^{\rm{H}}\left(\bm{A} \bm{x}^{(k-1)}-\bm y\right)\\
                {{\cal{X}}^{(k)}}:{\bm{x}}^{(k)} =S_\theta\left(\bm{z}^{(k)}\right)
                \end{array} \right. ,
\label{eq: ista-solver}
\end{equation}
where $k = 1,2,...,K$ denotes the ISTA iteration index;  $\rho = 1/L$ is the step size, where $L$ is usually taken as the largest eigenvalue of $ L =\bm A^T \bm A$; 
$S_{\theta}(\cdot)$ is usually a soft thresholding function corresponding to the sparse regularization of $\ell_1$-norm
\begin{align}
        S_\theta \left(\bm{x}^{(k)} \right)_i=\operatorname{sign}\left(x^{(k)}_{i}\right) \max \left(\left|x^{(k)}_{i}\right|-\theta,  0 \right),
\end{align}
where $\theta$ is the soft threshold, which is usually set as $\theta = \lambda \rho$.

Note that to obtain an accurate estimate $\bm {\hat{x}}$, ISTA usually requires hundreds of iterations. In addition, all the parameters such as $\rho$ and $\lambda$ are usually fixed, which requires careful tuning. To solve these problems, the UDNN corresponding to ISTA iterations in \eqref{eq: ista-solver} is proposed in the following subsections.

\subsection{Network Structure}

By taking full advantage of the merits of ISTA methods, the basic idea of UDNN is to map the previous ISTA iterations to a deep neural network that consists of a fixed number of layers, each of which corresponds to one iteration in ISTA. In particular, our UDNN can be summarized as the following three parts:

1) we map the ISTA iterations in \eqref{eq: ista-solver} to a
DNN graph, i.e., the $k$-th iteration of ISTA corresponds to the $k$-th layer of the DNN flow graph (see Fig. 1(a)). And the graph comprises two nodes respectively corresponding to two operations in \eqref{eq: ista-solver}, while arrows corresponding to the data flows between operations; 

2) In each layer of the graph, two types of nodes respectively map the nonlinear transform operation ${\cal{X}}^{(k)}$, defined by $S_{\theta}(\cdot )$, and the reconstruction operation ${\cal Z}^{(k)}$ in \eqref{eq: ista-solver} (see Fig. 1(b)).
The whole DNN flow graph is a multiple repetition of the above two parts corresponding to
successive iterations in ISTA. Given an under-sampled data, it flows over the graph
and finally generate the estimated $\bm{\hat{x}}$ in \eqref{eq:y2};

3) Two parallel propagation lines are used in the network to calculate the real and imaginary parts, respectively. Information is exchanged between the two propagation lines to ensure that the operation in the network is equivalent to the complex operation, while only real number is used (see Fig. 1(b)).

In this way, our deep UDNN retains the structure of ISTA, but generalizes the two operations (${{\cal{Z}}^{(k)}}$, ${{\cal{X}}^{(k)}}$) into two learnable layers, i.e., the reconstruction layer and nonlinear transform layer. Next, we present the details of these two layers.

\subsection{Reconstruction Layer}

In particular, let $\bm W_1 = \frac{1}{L}\bm A^{\rm{H}}$, $\bm W_2 = \bm I - \frac{1}{L}\bm A^{\rm{H}} \bm A$ and $\theta = \frac{1}{L}\lambda$. Then, for $k = 1,...,K$, given $\bm x^{(k-1)}$ and $\bm y$, the reconstruction layer of our UDNN is rewritten as follows (see Fig. 1(b)):
\begin{align}
  {\cal Z}^{(k)}: &~ \bm z^{(k)} = \left( \bm W_1^{(k-1)} \bm y + \bm W_2^{(k-1)} \bm x^{(k-1)} \right) \nonumber \\
&\Leftrightarrow \left\{ \begin{array}{l}
                \bm{z}_{\rm{R}}^{(k)} = {\bm{W}}^{(k)}_{2,{\rm{R}}} \bm {x}^{(k)}_{\rm{R}} - {\bm{W}}^{(k)}_{2,{\rm{I}}} \bm {x}^{(k)}_{\rm{I}}\\
                \;\;\;\;\;\;\;\;\;\; + {\bm{W}}^{(k)}_{1,{\rm{R}}} \bm{y}^{(k)}_{\rm{R}} - {\bm{W}}^{(k)}_{1,{\rm{I}}}\bm{y}^{(k)}_{\rm{I}}\\
                \bm{z}_{\rm{I}}^{(k)} = {\bm{W}}^{(k)}_{2,{\rm{R}}} \bm {x}^{(k)}_{\rm{I}} + {\bm{W}}^{(k)}_{2,{\rm{I}}} \bm {x}^{(k)}_{\rm{R}}\\
                \;\;\;\;\;\;\;\;\;\; + {\bm{W}}^{(k)}_{1,{\rm{R}}} \bm{y}^{(k)}_{\rm{I}} + {\bm{W}}^{(k)}_{1,{\rm{I}}} \bm{y}^{(k)}_{\rm{R}}
                \end{array} \right. .
  \label{eq: re-layer}
\end{align}
where $(\cdot)_{\rm{R}}$ and $(\cdot)_{\rm{I}}$ respectively denote the operator for taking the real and imaginary part. $\bm W_1^{(k-1)}$ and $\bm W_2^{(k-1)}$ are the learnable parameters, and $\bm y$ is the input data in \eqref{eq:y2}. Note that in the first layer, $\bm x^{(0)}$ is initialized to zero, hence, 
\begin{align}
        \bm z^{(1)} =  \bm W_1^k \bm y,
\end{align}
where $ \bm z^{(1)}$ can be regarded as an initial rough estimation of $\bm {x}$.

\subsection{Nonlinear Layer}
Nonlinear layer performs the nonlinear transform ${\cal X}^{(k)}$ in \eqref{eq: ista-solver}.
Given $\bm z^{(k)}$, the output of this layer is defined as 
\begin{align}
         x^{(n)}_i  =   \operatorname{sign}\left( z^{(n)}_i\right) \max \left(\left| z^{(n)}_i\right|-\theta^k, 0 \right), ~i=1,...,M,
\end{align}
where $x^{(n)}_i$ and $z^{(n)}_i$ are the $i$-th entry of $\bm x ^{(n)}$ and $\bm z ^{(n)}$, respectively, and $\theta^{(k)}$ is the learnable parameter in the $k$-th nonlinear transform layer.

        For the complex input $z^{(n)}_i$, we deal with real and imaginary parts of the input separately, i.e., $ x^{(n)}_i = S_\theta(z^{(n)}_{i,{\rm{R}}}) + jS_\theta(z^{(n)}_{i,{\rm{I}}}) $, where $j$ denotes the imaginary unit. That is, both real and imaginary parts share the same piecewise linear function.

\subsection{Network Parameters}
In the deep architecture of our UDNN, we aim to learn the following parameters: 
\begin{align}
        \varTheta &= \{ \bm W_1^{(k)},\bm W_2^{(k)}, \theta^{(k)}\}_{k=1}^{K} \nonumber  \\
 &= \{ \bm  W_{1,{\rm{R}}}^{(k)}, \bm W_{1,{\rm{I}}}^{(k)}, \bm W_{2,{\rm{R}}}^{(k)},\bm W_{2,{\rm{I}}}^{(k)}, \theta^{(k)}
                \}_{k=1}^{K}. 
\end{align}
Note that $\bm W_1^{(k)}$ and $ \bm W_2^{(k)}$ are all the parameters in the $k$-th reconstruction layer, and  $\theta^{(k)}$ is all the parameter in the $k$-th nonlinear transform layer. Hence, all parameters used for ISTA iteration are learnable in our network, which greatly increases the fitability of the network. Note that the real and imaginary parts of $ \bm W_1^{(k)}$ and $\bm W_2^{(k-1)}$ are separately used in the network, which requires a different training method that will be presented in the next section.

\section{UDNN Training}

\subsection{Training Data and Loss Function}

We first randomly generated $P$ sparse $\{\bm {x}_{p}\}_{p=1}^{P}$ as the training output data, then the measurement data $\{\bm {y}_{p}\}_{p=1}^{P}$ generated by \eqref{eq:y2} with $\{\bm {x}_{p}\}_{p=1}^{P}$ as the input is used as the training input data. Hence, the training set is $\Gamma = \{ \bm {y}_{p} , \bm {x}_{p} \}_{p=1}^{P} $. We choose the mean square error (MSE) as the training and testing loss function to train the network. Since the input and output of our UDNN are complex vectors, given pairs of training data, the loss between the network output and the ground-truth is defined as the sum of the real and imaginary parts:
\begin{align}
        E(\varTheta) = &~\frac{1}{|P|} \sum_{p = 1}^{P} \left\|  (\hat{\bm x}_p( \bm y_p, \varTheta))_{\rm{R}}- (\bm x_p)_{\rm{R}}\right\|_{2}^2, \\
         &~+ \frac{1}{|P|} \sum_{p = 1}^{P} \left\|  (\hat{\bm x}_p( \bm y_p, \varTheta))_{\rm{I}}- (\bm x_p)_{\rm{I}}\right\|_{2}^2,
\end{align}
where $\hat{\bm x}_p(\bm y_p, \varTheta)$ is the network output with parameter set $\varTheta$ and the $p$-th data $\bm y_p$ as the input.

\subsection{Initialization}

For deep neural networks, the parameters are generally initialized by random values. However, our UDNN is related with the ISTA iterations for solving \eqref{eq:y2}, we hence initialize the network parameters as the corresponding structures in ISTA, i.e., an initialized net is an exact realization of ISTA iterations for solving \eqref{eq: lasso}. Then, we train the network to improve the reconstruction accuracy.

\subsection{Gradient Computation}
We optimize the parameters of UDNN by using the gradient-based algorithm Adam. The gradients of loss function w.r.t. parameters are calculated by backpropagation over the deep architectures, which can be computed separately along the real and imaginary propagation lines. Alternatively, since only real number are used in the network, the gradients of UDNN hence can be calculated by using some popular deep learning toolboxes (e.g., the TensorFlow, PyTorch and Mat-ConvNet), as long as providing the proposed forward propagation in Fig.~1.

\begin{figure}[htbp]
\centerline{\includegraphics[width=3.8in]{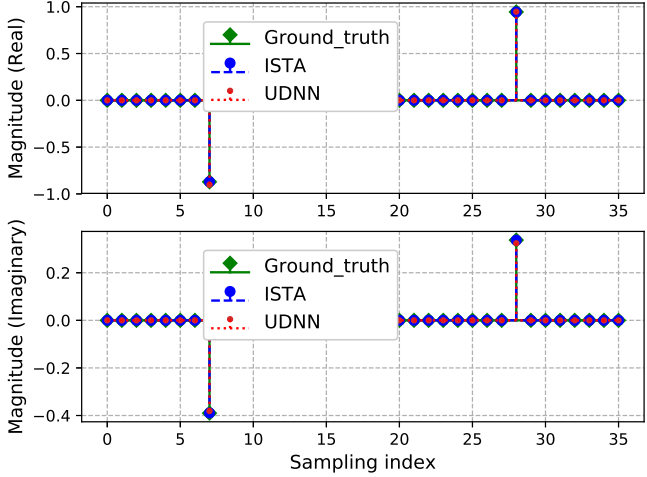}}
\caption{An example of the estimated $\bm{\hat{x}}$ under noiseless case. }
\label{figure:example}
\end{figure}

%
%

\section{Simulation Results}

We consider a single antenna OFDM channel estimation scenario with problem size of $16\times 36$, i.e., we set $N_d=N_b=4$ and $M = 36$ with $6$ frequency grid points are respectively generated in the delay and Doppler domain. In the simulation, we generate the ground-truths of delay and Doppler frequencies on the grid with a random location, and the real and imaginary part of the corresponding complex gains are randomly generated between $[-1,1]$. Note that this does not mean that our network can only handle signals with on-grid frequencies, this is just for facilitating the calculation of the validation loss used for comparison. Since we assume the data symbols are available, we set $\bm{\hat b}$ as all-one vector for simplicity. The SNR is defined according to \eqref{eq:ynm2} as $r/\sigma_w^2$ with $\sigma_w^2$ being the variance of the Gaussian noise samples. The number of layers of UDNN is set as $K = 5$. $L$ used in the ISTA and the network initialization is set as the largest eigenvalue of $ L =\bm A^T \bm A$. The weight factor $\lambda$ for the network initialization is set as $\lambda = 1$. We randomly generated 5 million training data to train the network, and let training go until the epoch number reaches $1000$. We implement UDNN by PyTorch, and the learning rate of Adam is set as $10^{-3}$ while other parameters of Adam are set as default. All experiments were conducted on a desktop computer with an Intel Core i7-9750H CPU and an RTX 2060 GPU.

\begin{figure}[htbp]
\centerline{\includegraphics[width=3.5in]{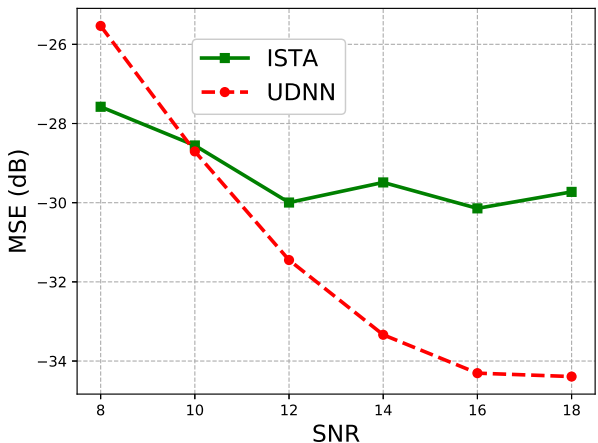}}
\caption{MSEs of $\bm x$ against SNRs. The results were averaged over 1000 runs. }
\label{figure:snr}
\end{figure}

%
%

We first present an example of the delay-Doppler estimation result under noiseless case in Fig.~\ref{figure:example}. The number of paths is set as $J = 2$. The weight factor $\lambda$ for ISTA is set as $\lambda = 0.01$. Once $\bm{\hat{x}}$ is obtained, the corresponding delay and Doppler frequencies can be easily determined by querying the corresponding value of the grid. From  Fig.~\ref{figure:example}  we can see that both UDNN and ISTA can obtain a good estimation result under noiseless cases. Next, we repeat this simulation 1000 times to compare the speed of ISTA and UDNN. The simulation results are presented in Table~\ref{tab1}, we can see that the average MSE of $\bm x$ obtained by ISTA over 1000 times is $-30.22$\,dB, the average number of iterations of ISTA is 2745, the running time of 1000 runs is $103.57$\,s. By comparison, the average MSE obtained by UDNN is $-35.38$\,dB,  since only $5$ layers are used in the network, the running time of 1000 runs is $0.21$\,s. The UDNN obtained a better performance than ISTA, while the computational complexity is significantly reduced.

\begin{table}[htbp]
\caption{Running Time Comparison}
\begin{tabular}{|c|c|c|c|}
\hline
\textbf{Methods} & \textbf{\textit{MSE}}& \textbf{\textit{Number of iterations/layers}}& \textbf{\textit{Running time}} \\
\hline
ISTA& $-30.22$\,dB & 2745 & $103.57$\,s  \\
\hline
UDNN& $-35.38$\,dB & 5 & $0.21$\,s  \\
\hline
\end{tabular}
\label{tab1}
\end{table}

Then, we compare the performance of ISTA and UDNN for noisy cases. The weight factor $\lambda$ for ISTA is set as $\lambda = \sigma_w \sqrt{2\log{(N)}}$. From Fig.~\ref{figure:snr} we can see that UDNN performs better than ISTA, except when SNR is very low. Because the training data of UDNN is generated under noiseless case. As for the condition of low SNR, we can increase the robustness of the network by adding noise to the training data. Of course, in return, the performance under high SNR will be reduced. We can choose whether to add noise to the training data according to the situation in practice.

\section{Conclusions}

In this paper, we concerned with the channel estimation problem of high mobility OFDM communication systems.  A two-dimensional compressed sensing problem, which considered the Doppler effect caused by the high mobility channel, was formulated by approximately linearizing the channel as a product of an overcomplete dictionary with a sparse vector. Then, we proposed an unfolded deep neural network for high mobility channel estimation, where the forward propagation is designed according to the ISTA. Two parallel propagation lines are used to respectively calculate the real and imaginary parts, and information are shared to ensure that the operation in the network with only real number is equivalent to the complex operation. Since all the parameters in UDNN are learnable, our UDNN can achieve better performance than ISTA with only few layers. Simulation results demonstrate that the proposed UDNN performs well for high mobility OFDM channel estimation while maintaining extremely fast computational speed. The future work will be to extend the UDNN to applications in massive MIMO configurations.

\bibliographystyle{IEEEtran}
\bibliography{mybible.bib}

\end{document}